\documentclass[]{piparticle-final}
\usepackage[T1]{fontenc}
\usepackage[utf8]{inputenc}
\usepackage{amsbsy}
\usepackage{float}
\usepackage{graphicx}
\usepackage{amsmath}

\providecommand{\tabularnewline}{\\}
\usepackage[american]{babel}

\usepackage{cite} 
\usepackage{epstopdf}

\begin{document}

\volume{5}               
\articlenumber{050002}   
\journalyear{2013}       
\editor{G. Mindlin}   
\received{9 October 2012}     
\accepted{1 March 2013}   
\runningauthor{M L Fern\'andez \itshape{et al.}}  
\doi{050002}         

\title{A mathematically assisted reconstruction of the initial focus of
the yellow fever outbreak in Buenos Aires (1871)}

\author{M L Fernández,\cite{inst1}
M Otero,\cite{inst2}
N Schweigmann,\cite{inst3}
H G Solari\cite{inst2}\thanks{E-mail: solari@df.uba.ar}
}

\pipabstract{
We discuss the historic mortality record corresponding to the initial
focus of the yellow fever epidemic outbreak registered in Buenos Aires
during the year 1871 as compared to simulations of a stochastic population
dynamics model. This model  incorporates the biology of the urban vector
of yellow fever, the mosquito \emph{Aedes aegypti}, the stages of
the disease in the human being as well as the spatial extension of the epidemic
outbreak. After introducing the historical context and the restrictions
it puts on initial conditions and ecological parameters, we discuss
the general features of the simulation and the dependence  on initial
conditions and available sites for breeding the vector. We discuss
the sensitivity, to the free parameters, of statistical estimators
such as: final death toll, day of the year when the outbreak reached
half the total mortality and the normalized daily mortality, showing
some striking regularities. The model is precise and accurate enough
to discuss the truthfulness of the presently accepted historic discussions
of the epidemic causes, showing that there are more likely scenarios
for the historic facts. 
}

\maketitle

\blfootnote{
\begin{theaffiliation}{99}
   \institution{inst1} Departamento de Computación, Facultad de Ciencias Exactas y Naturales (FCEN), Universidad de Buenos Aires (UBA) and CONICET. Intendente 
Güiraldes 2160, Ciudad Universitaria, C1428EGA Buenos Aires, Argentina.
   \institution{inst2} Departamento de Física, FCEN--UBA and IFIBA--CONICET. 1428 Buenos Aires, Argentina.
   \institution{inst3} Departamento de Ecología, Genética y Evolución, FCEN--UBA and IEGEBA--CONICET. 1428 Buenos Aires, Argentina.
\end{theaffiliation}
}

\section{Introduction}

Yellow fever (YF) is a disease produced by an \textbf{ar}thropod
\textbf{bo}rne \textbf{vi}rus (arbovirus) of the family \emph{flaviviridae}
and genus \emph{Flavivirus}. The arthropod vector can be one of
several mosquitoes and the usual hosts are monkeys and/or people.
Wild mosquitoes of genus \emph{Haemagogus}, \emph{Sabetes} and
\emph{Aedes} are responsible for the transmission of the virus among
wild monkeys, such as the Brown Howler Monkey (\emph{Alouatta guariba})
associated to recent outbreaks of YF in Brazil, Paraguay
and Argentina \cite{paho2008}). In contrast, urban YF is
transmitted by a domestic and anthropophilic mosquito, \emph{Aedes
aegypti},  human beings being the host \cite{chri60}. \emph{Aedes
aegypti} is a tree hole mosquito, with origins in Africa, that has
been dispersed through the world thanks to its association with people.

During the end of the XVIII and the XIX  centuries, YF caused
large urban outbreaks in the Americas from Boston (1798), New York
(1798) and Philadelphia (1793, 1797, 1798, 1799) in the North \cite{cart31}
to Montevideo (1857) and Buenos Aires (1858, 1870, 1871) \cite{penn95}
in the South. These historical episodes  arise as ideal cases for
testing the capabilities of YF models in urban settings.
Is it possible to reconstruct the evolution of one of these epidemic
outbreaks? Can enough information be recovered to produce a  thorough
test on the models? This is seldom the case, for example, for the
study of the Memphis (1878) epidemic, with over 10000 casualties, only
1965 were considered potentially usable \cite{arde03}. In contrast,
the records  of the outbreak in Buenos Aires 1871, unearthed and digitized for this work, left us with  an amount of 1274 death cases
located in time and space for the initial focus in the quarter of
San Telmo, about 78\% of the total mortality in the quarter \cite{acev73}.
According to the 1869 national census \cite{cens69} Buenos Aires
had 177787 inhabitants, 12329 of them living in San Telmo, about half
of them just immigrated into the country mostly from Europe.

In this work, we will compare the initial development of the epidemic
outbreak (Buenos Aires, 1871) with the simulations resulting from
an eco-epidemiological model developed in Refs. \cite{oter06,oter08,oter10},
 testing the worth of the predictive model.

The simulations were performed under a number of assumptions, most
of them essentially forced by the lack of better information. We will
assume  that: 
\begin{enumerate}
\item  Now and before, YF is the same illness, i.e., we can use
current information on YF development such as: the average
extent of the incubation,  infection, recovery, and toxic periods,
as well as the mortality level in 1871. In other words, the
virus presents no substantial changes since 1871 to present days.
We do not expect this hypothesis to be completely correct: the YF
virus is an \emph{RNA}-virus as opposed to the stable \emph{DNA}-viruses,
as such, mutations in about 140 years of continuous replications in
mosquitoes and primates can hardly be ruled out. Furthermore, present-day YF has been subject to different evolutionary pressures
than the YF in the XIX century. While in the XIX century yellow
fever circulated continuously in human populations, today the wild
part of the cycle involving wild populations of monkeys plays a substantial
role. 
\item The epidemic was transmitted by \emph{Aedes aegypti}. There is no
evidence of this fact since the scientific society and medical doctors
in general were not aware of the role played by the mosquito until
the confirmation given by Reed \cite{cart31} of Finlay's ideas \cite{finl99}.
\footnote{According to other sources, it was Beauperthuy \cite{agra08} the
first one to accurately describe the transmission of YF by mosquitoes,
as observed in the epidemic outbreak at Cumaná, Venezuela (1853), as
well as the efficient measures of protection taken by Native Americans,
the use of nets to prevent the spread of the epidemic \cite{tepa90}.%
}

We assume that \emph{Aedes aegypti} has not changed since then,
and/or there are no substantial changes in the life cycle, vector
capabilities and adaptation between the (assumed) population in 1871
and present-day populations in Buenos Aires city. After the eradication
campaign (1958--1965) \cite{sope65}, \emph{Aedes aegypti} was eradicated
from Buenos Aires \cite{mini64}. Hence, the present populations
result from a re-infestation and they are not the direct descendants of
the mosquitoes population of 1871.

\item Lacking time statistics for the duration of the different stages in
the development of the illness, reproduction of the virus and life
cycle of the mosquito, we use, as distribution for such events, a maximum
likelihood distribution subject to the constrain of the average value
for the cycle. In short, we use exponentially distributed times for
the next event for all type of events.
\item Finally, and most importantly, we assume that the human population mobility is not a factor in the local spread of the disease. We
anticipate one of our conclusions: this assumption is likely to be false for
the full development of the epidemic outbreak but seems reasonable for the
early (silent) development.  The study of the secondary foci of the epidemic
outbreak merits a detailed analysis of the social and political circumstances
related to human mobility and it is beyond the possibilities of this study.
\end{enumerate}

Since we want the test to be as demanding as possible, more information
is needed to simulate the outbreak eliminating sources of ambiguity
and parameters to be fitted using the same test data. We recovered
the following information: 
\begin{enumerate}
\item Estimations of daily temperatures. They are relevant since the
temperature regulates the developmental rates of the mosquitoes.
\item A very rough, anecdotal, estimation of the availability of breeding
sites (BS) that, ultimately, control the carrying capacity of the
environment, the number of vectors and the infection rate.
\item Human populations discriminated by block in the city.
\item Estimations of the date of arrival of the virus to the city, putting
bounds to the reasonable initial conditions for the simulation. 
\end{enumerate}
This climatological, social and historical information represents
a determining part of the reconstruction as it is integrated into
the model jointly with the entomological and medical information to
produce stochastic simulations of possible outbreaks to be compared
with the historic records of casualties.

We will show that the model predicts large probabilities for the occurrence
of YF in the given historical circumstances and it is also
able to answer why a minor outbreak in 1870 did not progress towards
a large epidemic. The total number of deaths and the time-evolution
of the death record will be shown to agree between the historical
record and the simulated episodes as well, within the original focus.

The rest of the manuscript will be organized as follows: we will begin
with the description of YF in section II, including 
the eco-epidemiological model. In section III, we
will address the relevant climatological, social and historical aspects.
In section IV, we will explore the sensitivity of the model
to initial conditions and the number of available breeding sites, discussing
the statistics more clearly influenced by vector abundance. The historic
mortality records and the simulated records are compared in section
\ref{Compara}. We will finally discuss the performance of the model in
section \ref{Discu}

\section{The disease}

\label{Enfer} We will simply quote the fact sheet provided by the
World Health Organization \cite{onu08} as the standardized description:
\begin{quote}
``YF is a viral disease, found in tropical regions of Africa
and the Americas. It principally affects humans and monkeys, and is
transmitted via the bite of \emph{Aedes} mosquitoes. It can produce
devastating outbreaks, which can be prevented and controlled by mass
vaccination campaigns.

The first symptoms of the disease usually appear 3–6 days after infection.
The first, or acute, phase is characterized by fever, muscle pain,
headache, shivers, loss of appetite, nausea and vomiting. After 3–4
days, most patients improve and symptoms disappear. However, in a
few cases, the disease enters a toxicphase: fever reappears, and
the patient develops jaundice and sometimes bleeding, with blood appearing
in the vomit (the typical \emph{vomito negro}).
About 50\% of patients who enter the toxic phase die within 10--14
days''.
\end{quote}
We add that the remission period lasts between 2 and 48 hours \cite{ops05},
and as it was mentioned in the introduction, not only \emph{Aedes} mosquitoes
transmit the disease.

\subsection{The model}

The yellow fever model is rather similar to the already presented
dengue model \cite{oter10}, the similarity corresponds to the fact
that dengue is produced by a \emph{Flavivirus} as well, it is transmitted
by the same vector and follows the same clinical sequence in the human being,
although with substantially lesser mortality.

The model describes the life cycle of the mosquito \cite{oter06}
and its dispersalafter a blood meal, seeking oviposition sites \cite{oter08}.
The mosquito goes through several stages: egg, larva, pupa, adult (non
parous), flyer (i.e., the mosquito dispersing) and adult (parous). In
each stage, the mosquito can die or continue the cycle with a transition
rate between the subpopulations that depends on the temperature. The
mortality in the larva stage is nonlinear and it regulates the population
as a function of the availability of breeding sites. Thus, the
transitions from adult to flyer are associated with blood meals, the
event that can transmit the virus from human to mosquito and vice-versa.
From the epidemiological point of view, the mosquito follows a SEI
sequence (Susceptible, Exposed ---extrinsic period---, Infective). Correspondingly,
the adult populations are subdivided according to their status with
respect to the virus. We assume that there is no vertical transmission
of the virus and eggs, larvae, pupae and non parous adults are always
susceptible.

The humans are subdivided in subpopulations according to their status
with respect to the illness as: susceptible (S), exposed (E), infective
(I), in remission (r), toxic (T) and recovered (R). The temporary
remission period is followed by recovery with a probability between
0.75 and 0.85 or a toxic period (probability 0.25 to 0.15) which ends
half of the times in death and half of the times in recovery. The
yellow fever model differs in the structure from the dengue model
in Ref. \cite{oter10}, as the human part of the dengue model
is SEIR and the yellow fever model is SEIrRTD. However, the additional stages
do not alter the evolution of the epidemic since the ``in
remission'', toxic and dead stages do not participate in the transmission
of the virus. The YF parameters are presented in Table \ref{parbo}.

\begin{table}[hbt]
\centering
\footnotesize
\begin{tabular}{lrr}
\hline \hline
Period  & value & range\tabularnewline
\hline 
Intrinsic Incubation (IIP) & 4 days & 3--6 days \tabularnewline
Extrinsic Incubation (EIP)  & 10 days & 9--12 days \tabularnewline
Human Viremic (VP)  & 4 days & 3--4 days \tabularnewline
Remission (rP)  & 1 days & 0--2 days \tabularnewline
Toxic (tP) & 8 days & 7--10 days \tabularnewline
\hline \hline
\end{tabular}
\vskip 1em
\begin{tabular}{lll}
\hline \hline
Probability & value  & range\tabularnewline
\hline 
Recovery after remission (rar)  & 0.75 & 0.75--0.85 \tabularnewline
Mortality for toxic patients (mt) & 0.5 & \tabularnewline
Transmission host to vector (ahv) & 0.75 & \tabularnewline
Transmission vector to host (avh) & 0.75 & \tabularnewline
\hline \hline
\end{tabular}
\caption{Parameters (mean value of state) adopted for YF.
The range indicated is taken from PAHO \cite{ops05}.\label{parbo}}
\end{table}

The model is compartmental, all populations are counted as non-negative
integers numbers and evolve by a stochastic process in which the time
of the next event is exponentially distributed and the events compete
with probabilities proportional to their rates in a process known
as density-dependent-Poisson-process \cite{durr01}. The model can
be understood qualitatively with the scheme of the Fig. \ref{esquema}.
The model equations are summarized in Appendix \ref{apendice}.

\begin{figure}[hbt]
\includegraphics[width=8cm]{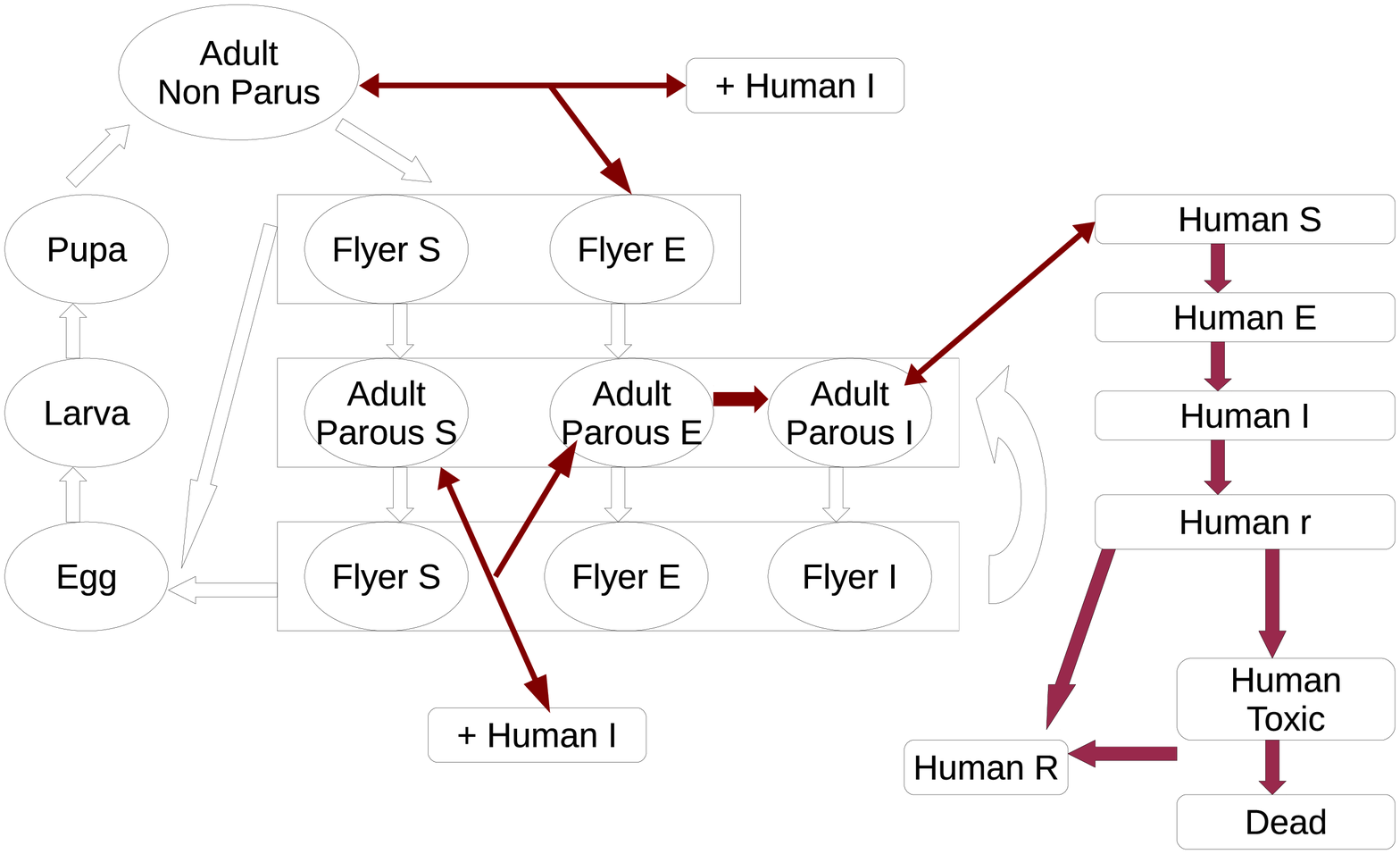}

\caption{Scheme of the yellow fever model. On the left side, the evolution of
the mosquito and, on the right side, the evolution of human subpopulations.
Hollow arrows indicate the progression through the life cycle of the
mosquito following the sequence: egg, larva, pupa, adult (non-parous),
flyer, adult (parous) and the repetition of the two last steps. The
mortality events are not shown to lighten the scheme. Eggs are laid
in the transition from flyer to adult. The adult mosquito populations
are subdivided according to their status with respect to the virus
as: susceptible (S), exposed (E) and infective (I). The virus is transmitted
from mosquitoes to humans and vice-versa in the transition from adult
to flyer (blood meal) when either the mosquito or the human is infective
and the other susceptible (red arrows). The red bold arrows indicate
the progression of the disease, from exposed to infective in the
mosquito and, in humans, following the sequence: exposed (E), infective (I),
in remission (r), toxic, recovered (R) or dead.\label{esquema}}
\end{figure}

The city is divided in blocks, roughly following the actual division (see Fig.
\ref{districtos}). The human populations are constrained to the block while the
mosquitoes can disperse from block to block.

\section{Historical, social and climatological information}

\label{Hist}

\subsection{When and how the epidemic started}

\begin{figure}[htb]
\includegraphics[width=7cm]{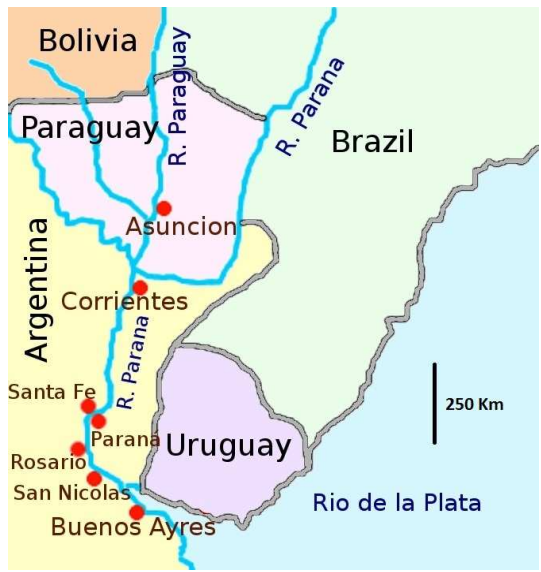}

\caption{A 1870 map \cite{mitc70} showing Asunción next to the Paraguay river,
Corrientes and Rosario next to the Paraná river and Buenos Aires (spelled
Buenos Ayres) next to the Rio de la Plata.\label{parana}}
\end{figure}

The YF outbreak in Buenos Aires (1871) was one of a series
of large epidemic outbreaks associated to the end of the War of the
Triple Alliance or Paraguayan war. The war confronted Argentina, Brazil
and Uruguay (the three allies) on one side and Paraguay on the other
side, and ended by March, 1870. By the end of 1870, Asunción, Paraguay's capital city, was  under the rule of the Triple Alliance. The return
of Paraguay's war prisoners from Brazil (where YF was almost
endemic at that time) to Asunción triggered a large epidemic outbreak
\cite{penn95}. The allied troops received their main logistic support
from Corrientes (Argentina), a city with 11218 inhabitants according
to the 1869 census \cite{cens69}, located about 300 km south of Asunción
(following the waterway) and 1000 km north of Buenos Aires along the
Paraná river (see Fig. \ref{parana}). On December 14, 1870, the
first case of YF was diagnosed in Corrientes \cite{more49},
and a focus developed around this case imported from Asunción. According
to some sources, the epidemic produced panic, resulting in about half
the population leaving the city between December 15 and January 15 
\cite{scen74}. However, other historical reasons might have played
a relevant role since the city of Corrientes was under the influence
and ruling of Buenos Aires, while in the farmlands, the General Ricardo
López Jordan was commanding a rebel army (a sequel of Argentina civil
wars and the war of the Triple Alliance). The subversion ended with
the battle of Ñaembé, about 200 km east of Corrientes, on January 26,
1871.

\begin{figure}[htb]
\includegraphics[width=8cm]{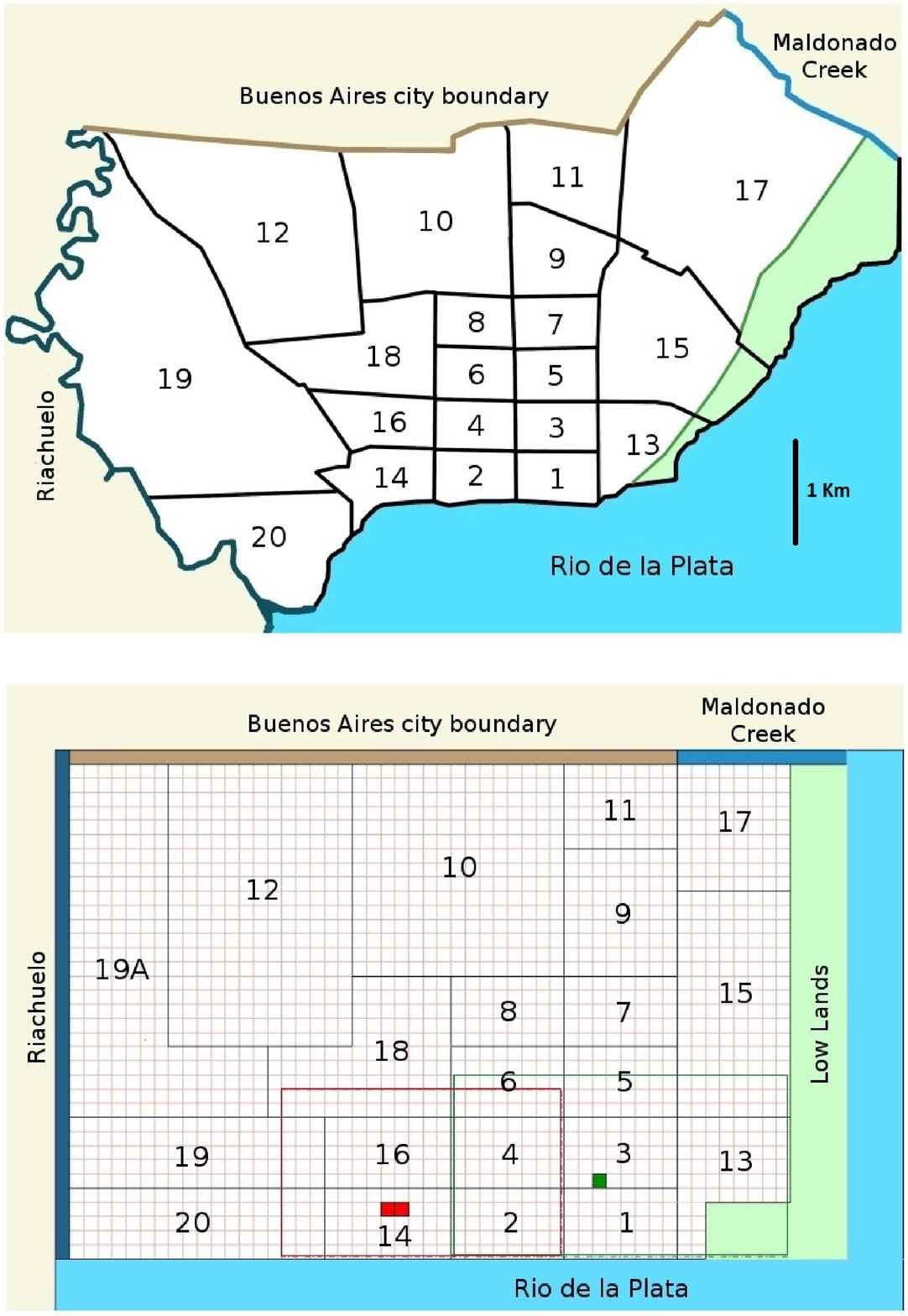}

\caption{Police districts from a map of the time and computer representation.
The red colored area in district 14 (San Telmo) are the two blocks where the 1871
epidemic started. The green colored area in district 3 is the block where the Hotel
Roma was and where the 1870 focus began (see section \ref{Compara}\ref{1870}).
The red and green lines sourround the region simulated for the 1871 epidemics
and the 1870 focus, respectively. Notice that districts 15
and 13 disagree in the maps. The computer representation follows the
information in Ref. \cite{cens89} from where the population information was
obtained. \label{districtos}}
\end{figure}

Putting things in perspective, we must realize that in those times, YF
was recognized only in its toxic stage associated to the black vomit, it is
then perfectly plausible that recently infected individuals would have left
Corrientes and Asunción reaching Buenos Aires, despite quarantine measures that
were late and leaky \cite{rmqu71,penn95}.\footnote{On December 16, a sanitary
official from Buenos Aires was commissioned to Corrientes to organize the
quarantine, a measure that was applied to ships coming from Paraguay but not to
those with Corrientes as departing port.%
} The death toll in Corrientes was of
1289 people in the city (and about 700 more in places around the city)
\cite{scen74}, representing a 11,5\% of the population (notice that this number
is not consistent with current numbers in use by WHO \cite{ops05} which
indicate a 7.5\% of mortality in diagnosed cases of YF but is well in
line with historical reports \cite{coop93} of 20\% to 70\% mortality in
diagnosed cases ---the statistical basis has changed with the
improved knowledge of early, not toxic, YF cases.

According to this historical view, the initial arrival of infectious
people to Buenos Aires happened, more likely, during December 1870 and
January 1871. In his study of the YF epidemic, written twenty
three years after the epidemic outbreak, José Penna (MD) \cite{penn95}
quotes the issue of the journal ``Revista Médico Quirúrgica'', published
in Buenos Aires on December 23, 1870 \cite{rmqu70}, which presents
a report regarding the sanitary situation during the last fifteen
days, indicating the emergence of a ``bilious fever'' and a general
tendency of other fevers to produce icterus or jaundice. In the next
issue, dated January 8, 1871, the ``Revista'' indicates an important
increase in the number of bilious fever cases reported \cite{rmqu71}.
In a separate article, the doctors call the attention on how
easily and how often the quarantine to ships coming from Paraguay is
avoided, and calls for strengthening the measures. Penna indicates
that the ``bilious fever'' (not a standard term in medicine)
likely corresponded to milder cases of YF. We will term this
idea ``Penna's conjecture'' and will come
back to it later.

For our initial guess, we considered this information as evidence
that the epidemic outbreak started during December, 1870. Exploring
the model, and arbitrarily, we took December 16, 1870, as the time
to introduce two infectious people with YF in the simulations,
at the blocks where the mortality started. Yet, an educated guess
for Penna's conjecture is to consider the 3--6 days needed from infection
to clinical manifestation and the 9--12 days of the extrinsic cycle. Hence, since the first clinical manifestations of transmitted YF
happened between December 11--23, we would guess the infected people
arriving somewhere between November 21 and December 11.

Yet, we must take into account that Penna's conjecture contrasts with
the conjectures presented by MD Wilde and MD Mallo, members of the
Sanity committee in charge during the YF epidemic. Wilde
and Mallo advocated for the spontaneous origin of the disease, very
much in line with the theories of miasmas in use in those times, theories
that guided the sanitary measures taken \cite{more49}.
Wilde and Mallo also argued that Asunción could not be the origin of
the epidemic, because of their belief that the ten or fifteen days
quarantine (counting since the last port touched) was enough to avoid
the propagation of the disease. This belief contrasts with the experience
of 1870 (in Buenos Aires) where a ten day quarantine was not enough
to prevent a minor epidemic \cite{penn95}. Nevertheless, the quarantine measures
werefully implemented in Corrientes by December 31,
1870. The measures were later lifted because of the epidemic in Corrientes
and implemented at ports down the Paraná river, being completed nearby
Buenos Aires (ports of La Conchas, Tigre, San Fernando and ``La
Boca'' within Buenos Aires city) by mid-February when
the epidemic was in full development in Buenos Aires according to
the port sanitary authorities, Wilde and Mallo \cite{more49}.

Being Corrientes the source of infected people can hardly be disregarded.
With about 5000 people leaving the city between December 15 and
January 15 \cite{more49}, a city where YF was developing.

According to Wilde and Mallo \cite{more49}, there were (non-fatal?)
YF cases in Buenos Aires as early as January 6, 1871, reported
by MDs Argerich and Gallarani as well as documented cases of YF
death after disembarking in Rosario (200 km North of Buenos Aires along
the Paraná river) having boarded in Corrientes.

\subsection{Breeding sites}

One of the key elements in the reconstruction and simulation of an
epidemic transmitted by mosquitoes is to have an estimation of their
numbers which will be reflected directly in the propagation of the
epidemic. In the mosquito model \cite{oter06}, this number is regulated
by the quality and abundance of breeding sites. The production of
a single breeding site, normalized to be a flower pot in a local cemetery,
was taken as unit in Ref. \cite{oter06} and the number of breeding sites
measured in this unit roughly corresponds to half a liter
of water.

The \emph{Aedes aegypti} population in monitored areas of Buenos
Aires, today, is compatible with about 20 to 30  breeding sites per block
\cite{oter08}. Estimating the number of sites available for breeding
today is already a difficult task, the estimation of breeding sites
available in 1871 is a nearly impossible one. In what remains of this
subsection, we will try to get a very rough a-priori estimate.

\begin{table}[hbt]
\centering
\footnotesize
\begin{tabular}{lcc}
\hline \hline
District  & Population/$(100\text{ m})^{2}$  & $\text{BS}/(100\text{ m})^{2}$\tabularnewline
\hline
1  & 339  & 391.0\tabularnewline
2  & 279  & 300.0\tabularnewline
3  & 428  & 522.0\tabularnewline
4  & 353  & 443.0\tabularnewline
5  & 330  & 430.0\tabularnewline
6  & 259  & 365.0\tabularnewline
13  & 160  & 196.0\tabularnewline
14  & 224  & 300.0\tabularnewline
15  & 90  & 157.0 \tabularnewline
16  & 165  & 316.0\tabularnewline
18  & 23  & 52.0\tabularnewline
19  & 13  & 26.0\tabularnewline
20  & 30  & 52.0\tabularnewline
\hline \hline
\end{tabular}

\caption{Population data. Buenos Aires, 1869 \cite{cens69}. Population density
by police district (see Fig. \ref{districtos}) and equivalent breeding sites,
$\text{BS}$, originally estimated as proportional to the house density in the police
district.
\label{poblacion}}
\end{table}

A very important difference between those days and the present corresponds
to the supply of fresh water which today is taken from the river,
processed and distributed through pipes; but in those days, it was
an expensive commodity taken from the river by the ``waterman''
and sold to the customers who, in turn, had to let it rest so that the
clay in suspension decanted to the bottom of the vessel (a process
that takes at least 3 days). Additionally, there were some wells available
but the water was (is) of low quality (salty). The last, and rather
common resource \cite{more49,herz79}, was the collection of rain
water in cisterns.

\begin{figure}[htb]
\includegraphics[height=8cm,angle=-90,clip]{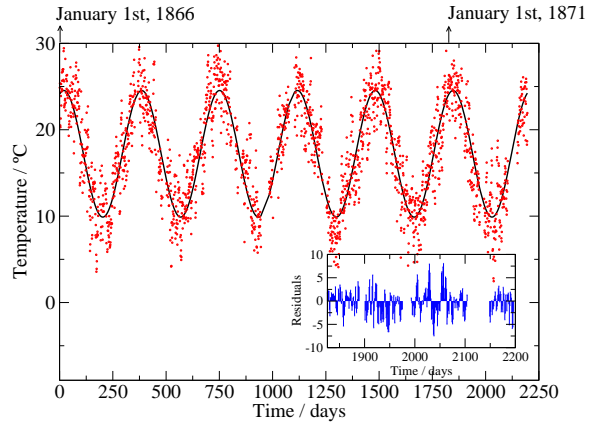}

\caption{Average daily temperature and periodic approximation fitted accordng
to Eq. (\ref{temperatura}), $t=0$ corresponds to January 1, 1866. The
inset shows the difference between the measured temperatures and the fit
(residuals) during 1871.  \label{fig-temp}}
\end{figure}

\subsection{Temperature reconstruction}

\emph{Aedes aegypti} developmental times depend on temperature.
Although it would seem reasonable to use as substitute of the real
data of the average temperature registered since systematic data collection
began, records of temperature in those times were kept privately \cite{goul78}
and are available. The data set consists of three daily measurements
made from January 1866 until December 1871, at 7AM, 2PM and 9PM. When
averaged, the records allow an estimation of the average temperature
of the day better than the usual procedure of adding maximum and minimum
dividing by two. Unfortunately, the register has some important missing
points during the epidemic outbreak. Because of this problem, the
data in Ref. \cite{goul78} was used to fit an approximation in the form:

\begin{align}
T=& 7.22^{\circ}\text{C} \times \cos(2\pi t/(365.25 \text{ days}) + 5.9484)\nonumber \\
& +17.21^{\circ}\text{C},\label{temperatura}
\end{align}
following Ref. \cite{kira02} and then extrapolating to the epidemic
period. In Fig. \ref{fig-temp}, the data and the fit are displayed.
The residuals of the fit do not present seasonality or sistematic
deviations, as we can see in the inset of Fig. \ref{fig-temp}.

It is worth noticing that a similar fit on temperature data from the
period 1980--1990 presents a mean temperature of $18.0^{\circ}$C, amplitude
of $6.7^{\circ}$C and a phase shift of $6.058^{\circ}$C \cite{oter06}
(notice that $t=0$ in the reference corresponds to July 1 while
in this work it corresponds to January 1). According to the threshold
computations in Ref. \cite{oter06}, the climatic situation was less favorable
for the mosquito in 1866--1871 than in the 1960--1991 period.

The reconstruction of temperatures needs to be performed at least from
the 1868 winter, since a relatively arbitrary initial condition in the form of
eggs for July 1, 1868 is used to initialize the code, and then run over a
transitory of two years. Such procedure has been found to give reliable results
\cite{oter06}. There are several factors in the biology of \emph{Ae.  ae.} that
indicate that the biological response to air-temperature fluctuations is
reflected in attenuated fluctuations of biological variables. First, the larvae
and pupae develop in water containers, thus, what matters is the water
temperature. This fact represents a first smoothing of air-temperature
fluctuations. Second, insects developmental rates for fluctuating temperature
environments correspond to averages in time of rates obtained in constant
temperature environments \cite{liu95}, an alternative view is that development
depends on accumulated heat \cite{fock93a}. Such averages occur over a period
of about 6 days at $30^\circ$C and longer times for other temperatures and
non-optimal food conditions \cite{gilp79}. Third, the biting rate (completion
of the gonotrophic cycle) depends as well on temperatures averaged over a
period of a few days. Last, mosquitoes actively seek the conditions that fit
best to them and more often than not, they are found resting inside the houses.

\subsection{Mortality data}

For this work, the daily mortality data recorded during the 1871
epidemic outbreak \cite{acev73} is key. This statistical work has received
no attention in the past, and no study of the YF outbreak in Buenos
Aires made reference to this information. We have cross-checked the
information with data in the 1869 national census \cite{cens69},
as well as with data in published works \cite{penn95}, and the details
are consistent among these sources.

The data set is presented here closing the historic research part.

\begin{figure}[htn]
\includegraphics[height=9cm,clip,angle=-90]{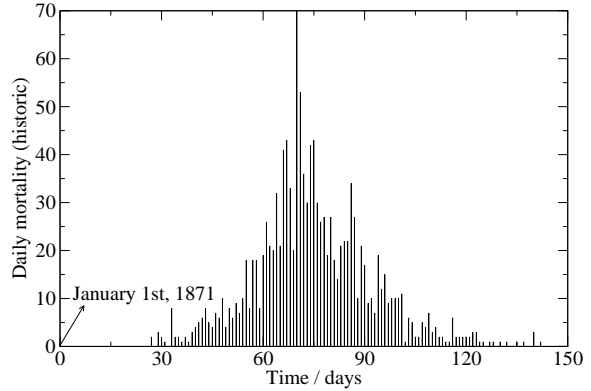}

\caption{Daily mortality cases in the police district 14 (see Fig. \ref{districtos})
corresponding to the quarter San Telmo \cite{acev73}, $t=0$ corresponds to January 1, 1871.\label{daily}} 
\end{figure}

\subsection{Revising the clinical development of yellow fever.}

A YF epidemic outbreak happened in Buenos Aires, in 1870, developing
about 200 cases \cite{penn95} (the text is ambiguous on whether
the cases are toxic or fatal). The epidemic outbreak was noticed by
February 22 (first death), a sailor who left Rio de Janeiro (Brazil)
on February 7, and presumably landed on February 17 (no cases
of YF were reported on board of the Poitou ---the boat). 

This well documented case allows us to see the margins of tolerance
that have to be exercised in taking medical information prepared for
clinical use as statistical information. Assume, following Penna,
that the sailor was exposed to YF before boarding in Rio
de Janeiro, according to information in Table \ref{parbo} collected
from the Pan American Health Organization \cite{ops05}, adding incubation
and viremic period, we have a range of 6--10 days, hence the sailor
was close to the limit of his infectious period. He was not toxic,
according to the MD on board who signed a certificate accepted by
the sanitary authority. Yet, five days later, he was dying, making
the remission plus toxic period of 5 days, shorter than the range
of 7--12 days listed in Table \ref{parbo} and substantially shorter
than the 10--15 days (remission plus toxic) communicated in Ref. \cite{onu08}.
Shall we assume as precise the values reported in Ref. \cite{onu08,ops05}
we would have to conclude that the disease was substantially different,
at least in its clinical evolution, in 1871 as compared with present
days.

In clinical studies performed during a YF epidemic on the
Jos Plateau, Nigeria, Jones and Wilson \cite{jone72} report a 45.6\%
overall mortality and a significant difference in the duration of
the illness for fatal and non-fatal cases with averages of 6.4 and
17.8 days. Série \emph{et al.} \cite{seri68a} reports for the 1960--1962
epidemic in Ethiopia a mortality ranging from 43\% (Kouré) to 100\%
(Boloso) and 50\% (Menéra) with a total duration of the clinic phase
of the illness of 7.14, 2.14 and 4.5 days, respectively (weighted average
of 4.6 days in 18 cases).

We must conclude that the extension of the toxic period preceding
the death presents high variability. This variability may represent
variability in the illness or in medical criteria. For example, Série
\cite{seri68a} indicates that the 100\% mortality found at the Boloso
Hospital is associated to the admission criteria giving priority to
the most severe cases. In correspondence with this extremely high
mortality level, the survival period is the shortest registered. The minimum
length of the clinical phase is of 10--14 days or 13--18 days, depending
of the source (adding viremic, remission and toxic periods). We note
that not only the toxic period of fatal cases must be shorter than
the same period for non-fatal cases, but also the viremic period must
be shorter in average, if all the pieces of data are consistent.

The time elapsed between the first symptoms and death is probably
longer today than in 1871, since it, in part, reflects the evolution
of medical knowledge. The hospitalization time is also rather arbitrary
and changes with medical practices which do not reflect changes in
the disease.

A rudimentary procedure to correct for this differences is to shift the
simulated mortality some fixed time between 5 and 8 days (the difference
between our 13 days guessed (Table \ref{parbo}) and the 4.5--6.4 reported for
Africa \cite{seri68a,jone72}). Such a procedure is not conceptually optimal,
but it is as much as it can be done within present knowledge. We certainly do not
know whether just the toxic period must be shortened or the viremic period must
be shortened as well, and in the latter case, how this would affect the
spreading of the disease.

A second source of discrepancies between recorded data and simulations are the
inaccuracies in the historic record. Can we consider the daily mortality record
as a perfect account? Which was the protocol used to produce it? We can hardly
expect it to be perfect, although we will not make any provision for this
potential source of error.

\section{Simulation results}

\label{Model2} The simulations were performed using a one-block spatial
resolution, with the division in square blocks of the police districts 14 (San
Telmo), 16, 2 and 4, corresponding to Concepción, Catedral Sur and Montserrat;
and part of the districts 6, 18, 19, 19A and 20 (see Fig. \ref{districtos}),
and totalized for each police district to obtain daily mortality
comparable to those reported in Ref. \cite{acev73} and picture in Fig.
\ref{daily}. Numerical mosquitoes were not allowed to fly over the river. At
the remaining borders of the simulated region, a Stochastic Newmann Boundary
Condition was used, meaning that the mosquito population of the next block
across the boundary was considered equal to the block inside the region; but the
number of mosquito dispersion events associated to the outside block was drawn
randomly, independently of the events in the corresponding inner block. Larger
regions for the simulations were tested producing no visible differences.

The time step was set to the small value of $30$ s, avoiding the introduction of further complications in the program related to fast event rates for
tiny populations \cite{sola03}, although an implementation of the method in
Ref. \cite{sola03}, not relying on the smallness of the time step so heavily, is
desirable for a production phase of the program.

Before we proceed to the comparison between the historic mortality
records of the epidemic and the simulated results, we need to gain
some understanding regarding the sensitivity of the simulations to
the parameters guessed and the best forms of presenting these results.
We performed a moderate set of computations, since the code has not
been optimized for speed and it is highly demanding for the personal
computers where it runs for several days. Here, we illustrate the
main lessons learned in our explorations. Poor people, unable
to buy large quantities of water, had to rely mostly on the cisterns
and other forms of keeping rain water. Since 1852, when the population
of Buenos Aires was about 76000 people, there was an important immigration
flow, increasing the population to about 178000 people by 1869 \cite{cens69}.
The immigrants occupied large houses where they rented a room, usually
for an entire family, a housing that was known as ``conventillo''
and was the dominant form of housing in some districts such as San
Telmo, where the epidemic started \cite{scen74}. In some police
chronicles of the time, houses with as many as 300 residents are mentioned
\cite{bare80}. Under such difficult social circumstances, we can
only imagine that the number of breeding sites available to mosquitoes
has to be counted as orders of magnitude larger than present-day
available sites.

An a-priori and conservative estimation is to consider about ten times
the number of breeding sites estimated today. Thus, we assume, as a
first guess, 300 breeding sites per block in San Telmo. We will have
to tune this number later as it regulates mosquito populations and
the development of the epidemic focus. The number of breeding sites
is the only parameter tuned to the results in this work.

More precisely, the criteria adopted was to make the number of (normalized)
BS proportional to the number of houses per block taken from historic
records \cite{cens89}, adjusting the proportionality factor to the
observed dynamics. We introduce the notation BSxY to indicate
a multiplicative factor of Y.

The population of each police district was set to the density values
reported in the 1869 census \cite{cens69,cens89} and the police
districts geography was taken from police records \cite{roma66}
and referenced according to maps of the city at the time \cite{bsas87,mitr80,solv63,mitr62}.

Table {\ref{poblacion}} shows the average population per block,
initially estimated number of breeding sites and number of
houses for the district of the initial focus, San Telmo \#14 and nearby
districts (\# 16, 4, 2). A sketch of Buenos Aires police districts
according to a 1887 map \cite{bsas87} is displayed alongside with
the computer representation in Fig. \ref{districtos}.

It is a known feature of stochastic epidemic models \cite{ande00} that the
distribution of totals of infected people has two main contributions. One is that the
small epidemic outbreaks when none or a few secondary cases are produced and the
extinction time of the outbreak comes quickly. The otheris that the large epidemic
outbreaks which, if the basic reproductive number is large enough, present a
Gaussian shape separated by a valley of improbable epidemic sizes from the
small outbreaks.

While the present model does not fall within the class of models discussed
in Ref. \cite{ande00}, the general considerations applied to stochastic
SIR models qualitatively apply to the present study. Yet, simulations
started early during the summer season follow the pattern just described in
Ref. \cite{ande00}, but simulations started later do not present the probability
valley between large and small epidemics.

We have found useful to present the results disaggregated in the form:
epidemic size, daily percentage of mortality relative to the total
mortality and time to achieve half of the final mortality. This presentation
will let us realize that most of the fluctuation is concentrated in
the total epidemic size, while the daily evolution is relatively regular,
except, perhaps, in the time taken to develop up to 50\% of the mortality
(depending on the abundance of vectors and the initial number of infected
humans and chance).

\subsection{Total mortality (epidemic size)}

Since historical records include mostly the number of causalities,
it appears sensible for the purposes of this study to use the total
number of deaths as a proxy statistics for epidemic size.

\begin{figure}[htb]
\includegraphics[height=9cm,angle=-90]{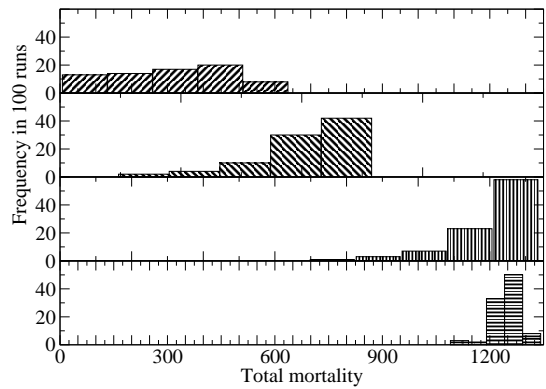}

\caption{San Telmo. Total mortality histograms for different number of breeding
sites, computed after 100 simulations with the same initial condition
corresponding to 2 infectious people located in San Telmo on January
1, 1871, at the same location where the initial death happened in
the historical event. From top to bottom, multiplication factors (bin-width
: frequency of no-epidemic) BSx1 (126.2 : 0.23), BSx2 (209.8 : 0.09),
BSx3 (128.2 : 0.06) and BSx4 (50.6 : 0.02). The y-axis indicates frequency
in a set of 100 simulations.\label{muerte-bs}}
\end{figure}

The total mortality depends strongly on the stochastic nature of the
simulations, initial conditions and ecological parameters guessed.
Qualitatively, the results agree with the intuition, although this
is an a-posteriori statement, i.e., only after seeing the results
we can find intuitive interpretations for them.

The discussion assumes that the development of the epidemic outbreak
was regulated by either the availability of vectors (mosquitoes) or
the exhaustion of susceptible people, the first situation represents
a striking difference with standard SIR models without seasonal dependence
of the biological parameters. Actually, in Fig. \ref{muerte-bs},
we compare frequencies of epidemics binned in five bins by final epidemic
size for different sets of 100 simulations with different number of
breeding sites. The number of breeding sites is varied in the same
form all along the city, keeping the proportionality with housing,
and it is expressed as multiplicative factor (BSx1=1, BSx2=2, BSx3=3,
BSx4=4) presented in Table \ref{poblacion}. Notice also that the
width of the bins progresses as 126.2, 209.8, 128.2 and 50.6 indicating
how the dispersion of final epidemic sizes first increases with the
number of breeding sites but for larger numbers decreases.

We can see how for a factor 2 (BSx2) (and larger) the mortality saturates,
indicating the epidemic outbreaks are limited by the number of susceptible
people. For our original guess, factor 1, the epidemic is limited
by the seasonal presence/absence of vectors. However, for higher factors,
there is a substantial increase in large epidemic outbreaks with larger
probabilities for larger epidemics. Only for the factor 4 (BSx4) the
most likely bin includes the historical value of 1274 deaths.

\begin{figure}[htb]
\includegraphics[height=9cm,clip,angle=-90]{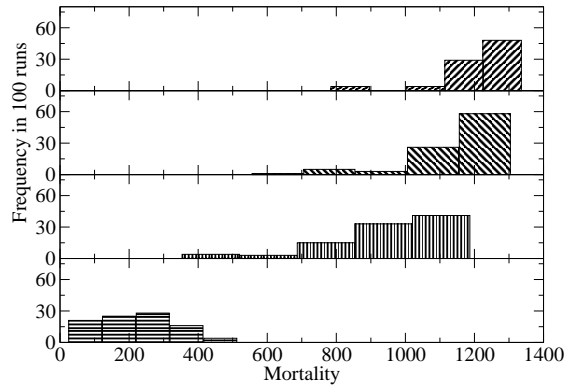}

\caption{San Telmo. Total mortality histograms for different date of arrival
of infectious people. Computed after 100 simulations with the same
initial condition corresponding to 2 infectious people located in
San Telmo and a density factor BSx2.5. From top to bottom: December
16, 1870; January 1, January 16 and February 15, 1871.\label{muerte-fecha}}
\end{figure}

A second feature, already shown in the Dengue model \cite{oter10},
is that outbreaks starting with the arrival of infectious people in
late spring will have a lesser chance to evolve into a major epidemic.
Yet, those that by chance develop are likely to become large epidemic
outbreaks since they have more time to evolve. On the contrary,
outbreaks started in autumn will have low chances to evolve and not
a large number of casualties. The corresponding histograms can be
seen in Fig. \ref{muerte-fecha}. Figure \ref{muerte-fecha} also
shows how the outbreaks that begin on December 16, as well as simulations
starting on January 1, present higher probabilities of large epidemics
than of small epidemics, but this tendency is reverted in simulations
of outbreaks that start by February 16. This transition is, again,
the transition between outbreaks regulated by the number of available
susceptible humans and those regulated by the presence or absence
of vectors.

\subsection{Mortality progression}

\begin{figure}[htb]
\includegraphics[height=9cm,clip,angle=-90]{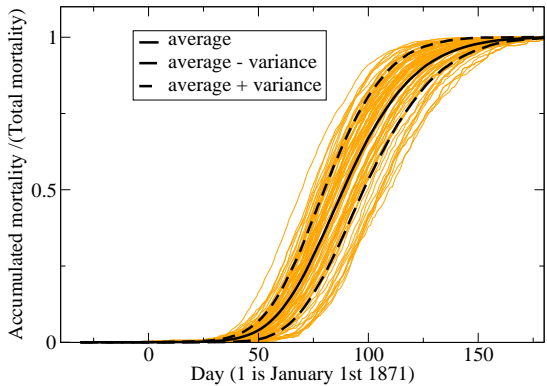}

\caption{Evolution of the mortality level for all the epidemics with three or more
fatal cases in a batch of 100 simulations (96 runs) differing only
in the pseudo-random series. The breeding sites number has a factor
4, and the initial contagious people were placed on January 1, 1871,
in San Telmo.\label{colapso}}
\end{figure}

One of the most remarkable facts unveiled by the simulations is that
when the time evolution of the mortality is studied as a fraction
of the total mortality, much of the stochastic fluctuations are eliminated
and the curves present only small differences (see Fig. \ref{colapso}). The similarity of the normalized evolution allows us to focus on
the time taken to produce half of the mortality (labelled $T_{1/2}$).

We notice, in Fig. \ref{colapso}, that the $T_{1/2}$ in these runs
lay between 69--110, compared to the historic values of $T_{1/2}=$73.

\begin{figure}[hbt]
\includegraphics[height=9cm,clip,angle=-90]{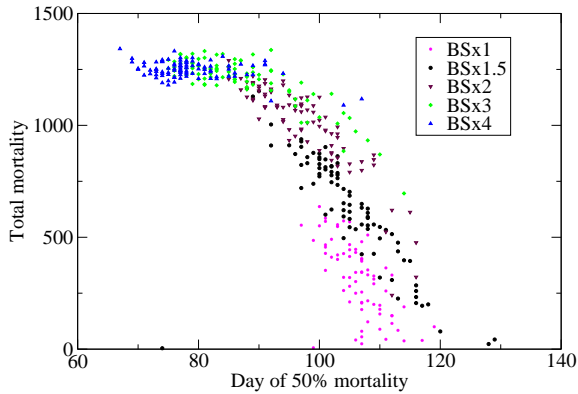}

\caption{San Telmo. Total mortality versus day when the mortality reached half
its final number for different number of breeding sites. The mortality
appears to be mainly a function of $T_{1/2}$ and roughly independent
of the number of breeding sites.\label{TvsZ}}
\end{figure}

The latter observation brings the attention to a remarkable fact of
the simulations: not only the normalized progression of the outbreaks
are rather similar but also, there is a correspondence between early
development and large mortality. Drawing $T_{1/2}$ against total-mortality
(Fig. \ref{TvsZ}), we see that even for different number of breeding
sites, all the simulations indicate that the final size is a noisy
function of the day when the mortality reaches half its final value.
This function is almost constant for small $T_{1/2}$ and becomes
linear with increasing dispersion when $T_{1/2}$ is relatively large.
Once again, the two different forms the outbreak is controlled.

\section{Real against simulated epidemic}
\label{Compara}

We would like to establish the credibility of the statement: the historic
mortality record for the San Telmo focus belongs to the statistics
generated by the simulations.

To achieve this goal, we need to compare the daily mortality in the
historical record and the simulations. Since, in the model, the mortality
proceeds day after day with independent random increments (as a consequence
of the Poisson character of the model), it is reasonable to consider
the statistics 
\begin{equation}
\chi^{2}=\sum_{i}\left(\frac{HM(i)-MM(i)}{D(i)}\right)^{2}\label{chi2def}
\end{equation}
 where $i$ runs over the days of the year, $HM(i)$ is the fraction
of the total death toll in the historic record for the day $i$, $MM(i)$
is the average of the same fraction obtained in the simulations and
$D(i)$ is the corresponding standard deviation for the simulations.
The sum runs over the number of days in which the variance is not
zero, for BSx3 and BSx4 in no case $D(i)=0$ and $(HM(i)-MM(i)\ne0 )$.
The number of degrees corresponds to the number of days with non-zero
mortality in the simulations minus one. The degree discounted accounts
for the fact that $\sum_{i}HM(i)=\sum_{i}MM(i)=1$.

\subsection{Tuning of the simulations}

Before we proceed, we have to find an acceptable number of breeding
sites, a reasonable day for the arrival of infected individuals (assumed
to be 2 individuals arbitrarily) and adjust for the uncertainty in
survival time. Actually, moving the day of arrival $d$ days earlier
and shortening the survival time by $d$ will have essentially the
same effect on the simulated mortality (providing $d$ is small), which
is to shift the full series by $d$ days. This is, assigning to the
day $i$ the simulated mortality $SM$ of the day $i+d$, ($SM(i+d)$).
Hence, only two of the parameters will be obtained from this data.

\begin{table}[hbt]
\centering
\begin{tabular}{ccccc}
\hline \hline
BSxY  & d  & $\chi^{2}$  & degree  & Probability\tabularnewline
\hline
2  & 8  & 490.2  & 177  & 0. \tabularnewline
3  & 7  & 191.7  & 173  & 0.15 \tabularnewline
4  & 3  & 151.9  & 168  & 0.80 \tabularnewline
4  & 4  & 143.8  & 168  & 0.91 \tabularnewline
4  & 5  & 145.6  & 168  & 0.89 \tabularnewline
\hline \hline
\end{tabular}

\caption{$\chi^{2}$ calculations according to (\ref{chi2def}). We indicate
the multiplicative factor applied to the breeding sites described
in Table \ref{poblacion}, the shift applied to the statistics produce
with the parameters of Table \ref{parbo}, the value of the statistic $\chi^2$,
 the number of degrees of
freedom, and the probability $P(x>\chi^{2})$ for a random variable
$x$ distributed as a $\chi^{2}$-distribution with the indicated
degrees of freedom. The standard reading of the $\chi^{2}$ test indicates
that, for BSx3 and BSx4, the statement ``the deviations from
the simulations mean of the historic record (deviations assumed to
be distributed as $\chi^{2}$ with the indicated degrees) belong to
the simulated set'' is not found likely to be false by
the test in the cases BSx3 and BSx4.\label{chi2}}
\end{table}

As we have previously observed, the total mortality presents a large variance
in the simulations. Moreover, in medical accounts of modern time \cite{jone72,seri68a},
the mortality ranges between $46\%$ and $100\%$ while in historical
accounts the percentage goes from $20$ to $70$ \cite{coop93}.
Hence, a simple adjustment of the mortality coefficient from our arbitrary
$50\%$ within such a wide range would suffice to eliminate the contributions
of the total epidemic size. The average simulated epidemic for BSx4
is of $\approx 1248$ deaths while the historic record is of $1274$ deaths. Hence,
to match the mean with the historic record, it would suffice to correct
the mortality from $50\%$ to $51\%$.

We can disregard the idea that the epidemic started before December
20, 1870 (Penna's conjecture), since it is not possible to simultaneously obtain
an acceptable final mortality and an acceptable evolution
of the outbreak. Our best attempt corresponds to an epidemic starting
by December 14, 1870, which averages $\approx 1212$ deaths and with a deviation
of the mortality of $\chi^{2}=219.1$ with 175 degrees, giving a probability
$P(x\ge219,1)=0.01$.

We focus on arrival dates around January 1, 1871. We can also disregard,
for this initial condition, the original guess BSx1 corresponding to
300 breeding sites per block in San Telmo, since it produces too small
epidemics. We present results for the epidemics corresponding to BSx2,
BSx3 and BSx4 in Table \ref{chi2} for different numbers of BS and
shift $d$.

\subsection{Comparison}

\begin{figure}[htb]
\includegraphics[viewport=0bp 0bp 565bp 842bp,clip,angle=-90,width=9cm]{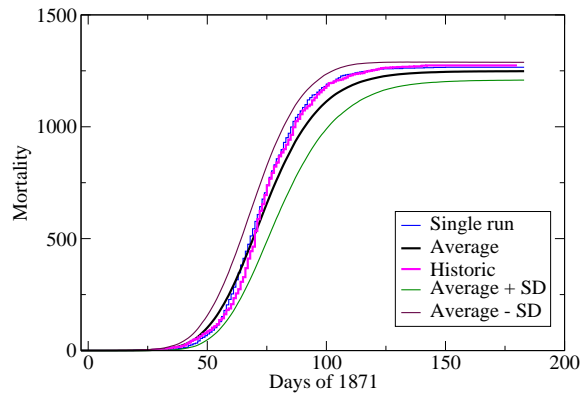}

\caption{The historic record of accumulated mortality as a function of the
day of the year 1871 is presented. The averaged accumulated mortality
as well as curves shifted one standard deviation are shown for comparison.
An acceptable look alike individual simulation chose by visual inspection
from a set of 100 simulations is included as well. Two exposed individuals
(not yet contagious) were introduced in the same blocks where the
historic epidemic started the day January 1, 1871, with BSx4. The
simulated mortality is anticipated in 4 days. Statistical estimates
were taken as averages over 96 runs which resulted in secondary mortality,
out of a set of 100 runs.\label{compare}}
\end{figure}

The conclusion of the $\chi^{2}$ tests is that the simulations performed
with BSx3 and beginning between December 24 1870 and January 1,
1871 (with a survival period shortened between 0 and 8 days) are compatible
with the historical record. However, the compatibility is larger when
BSx4 is considered and the beginning of the epidemic is situated between
December 28, 1870 and January 5, 1871 (with a survival period shortened
between 0 and 8 days respectively). We illustrate this comparison
with Fig. \ref{compare}.

\subsection{The 1870 outbreak}
\label{1870}

The records for the 1870 outbreak are scarce. Of the recognized cases,
only 32 entered the Lazareto (hospital) and 19 of them were originated
in the same block that the first case. Secondary cases are registered
at the Lazareto' books starting on March 30 (two cases) and continuing
with daily cases. the final outcome of these cases and the cases not entered at the Lazareto \cite{penn95} are not clear. 

Except for the precise initial condition, corresponding to one viremic
(infective) person located at the Hotel Roma (district 4 in Fig.
\ref{districtos}), the information is too imprecise to produce a
demanding test for the model.

We performed a set of 100 simulations introducing one viremic (infective)
person at the precise block where the Hotel Roma was, on February
17. The number of breeding sites was kept at the same factor 4 with
respect to the values tabulated in Table \ref{poblacion} that was used
for the best results in the study of the 1871 outbreak. Needless to
say, this does not need to be true, as the number of breeding sites may
change from season to season.

\begin{figure}[htb]
\includegraphics[height=9cm,clip,angle=-90]{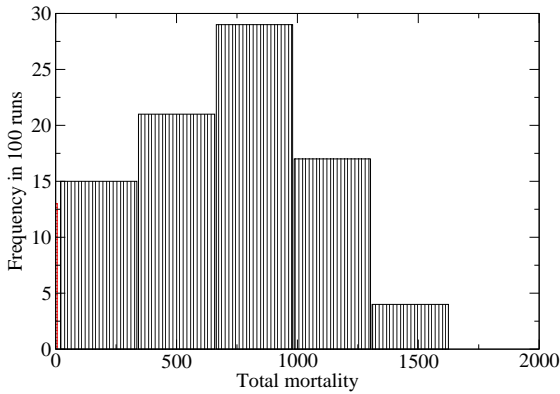}

\caption{Mortality distribution for the simulations beginning on February 17
with the incorporation of one infective (viremic) in the block of
the Hotel Roma and BSx4. The histogram is the result of 87 runs
which resulted in epidemics (13 runs did not result in epidemics). The
width of the bins is 322.2, and the first epidemic bin goes from 17
to 339 deaths with a frequency of 15/100. \label{H1870}}
\end{figure}

The distribution of the final mortality is shown in Fig. \ref{H1870}.
As we can see, relatively small epidemics of less than 200 deaths
cannot be ruled out, although there are much larger epidemics also
likely to happen.

\begin{figure}[htb]
\includegraphics[height=9cm,clip,angle=-90]{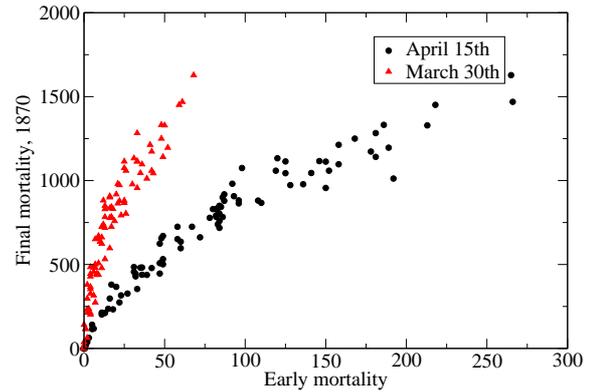}

\caption{Final mortality against early mortality for two dates: March 30 and
April 15. A low early mortality ``predicts'' a low final mortality
as the outbreak does not have enough time to develop. Simulations
correspond to the conditions of the 1870 small epidemic. BSx4, one
infected arriving to the Hotel Roma on February 17. The historic
information indicates that secondary cases were recorded by March 30.
Hence, corresponding to a slow start. The final mortality is not known,
but it is believed it has been in the 100--200 range.\label{abril15}}
\end{figure}

Actually, a slow start of the epidemic outbreak would favor a small
final mortality, as it can be seen in Fig. \ref{abril15}. Not only
there is a relation between a low mortality early during the outbreak
(such as April 15) and the final mortality, but we also see that
the sharp division between small and large epidemics is not present
in this family of epidemic outbreaks differing only in the pseudo-random
number sequence.

For the smallest epidemics simulated, the secondary mortality starts
after March 30. Hence, the 1870 focus can be understood as a case
of relative good luck and a late start within more or less the same
conditions than the outbreak of 1871.

\section{Conclusions and final discussion}

\label{Discu} In this work, we have studied the development of the
initial focus in the YF epidemic that devastated Buenos Aires
(Argentina) in 1871 using methods that belong to complex systems epistemology
\cite{garc07}.

The core of the research performed has been the development of a model
(theory) for an epidemic outbreak spread only by the mosquito \emph{Aedes
aegypti} represented according to current biological literature such
as Christophers \cite{chri60} and others. The translation of the
mosquito's biology into a computer code has been performed earlier
\cite{oter06,oter08} and the basis for the spreading of a disease
by this vector has been elaborated in the case of Dengue previously
\cite{oter10}. The present model for YF is then an adaptation
of the Dengue model to the particularities of YF, and the
present attempt of validation (failed falsification) reflects also on
the validity of this earlier work.

The present study owes its existence to the work of anonymous police
officers \cite{acev73} that gathered and recorded epidemics statistics
during the epidemic outbreak, in a city that was not only devastated
by the epidemic, but where the political authorities left in the middle
of the drama as well \cite{scen74}.

We have gathered (and implemented in a model) entomological, ecological
and medical information, as well as geographic, climatological and
social information. After establishing the historical constraints
restricting our attempts to simulate the historical event, we have
adjusted the density of breeding sites to be the equivalent to 1200
half-liter pots as those encountered in today Buenos Aires cemeteries
(the number corresponds to San Telmo quarter). Perhaps a better idea
of the number of mosquitoes present is given by the maximum of the
average number of bites per person per day estimated by the model,
which results in 5 bites/(person day) (to be precise: the ratio between
the maximum number of bites in a block during a week and the population
of the block, divided by seven).

The population of the domestic mosquito \emph{Aedes aegypti} in
Buenos Aires 1870--1871 was large enough to almost assure the propagation
of YF during the summer season. The only effective measures
preventing the epidemic were the natural quarantine resulting from
the distance to tropical cities were YF was endemic (such
as Rio de Janeiro) and the relatively small window for large epidemics,
since the extinction of the adult form of the mosquito during the
winter months prevents the overwintering of YF. In this sense,
the relatively small outbreak of 1870 is an example of how a late
arrival of the infected individual combined with a touch of luck produced
only a minor sanitary catastrophe.

By 1871, as a consequence of the end of the Paraguayan war and the
emergence of YF in Asunción, the conditions for an almost
unavoidable epidemic in Buenos Aires were given. The intermediate step
taken in Corrientes, with the panic and partial evacuation of the
city, adding the lack of quarantine measures, was more than enough
to make certain the epidemic in Buenos Aires. On the contrary, Penna's
conjecture of an earlier starting during December, 1870 are inconsistent
with the biological and medical times as implemented in the model.
We can disregard this conjecture as highly improbable.

The historic mortality record is consistent with an epidemic starting
between December 28 and January 5, being the symptomatic period
(viremic plus remission plus toxic) of the illness between 13 and
5 days. Furthermore, the existence of non-fatal cases of YF
by January 6 mentioned by some sources \cite{more49} would be
consistent, provided the cases were imported.

In retrospect, the present research began as an attempt to validate/falsificate
the YF model and, in more general terms, the model for the transmission
of viral diseases by \emph{Aedes aegypti} using the historic data of this
large YF epidemic. As the research progressed, it became increasingly
evident that the model was robust.  In successive attempts, every time the
model failed to produce a reasonable result, it forced us to revise the
epidemiological and historical hypotheses. In these revisions, we ended up
realizing that the accepted origin of the epidemic in imported cases from
Brazil, actually hides the central role that the epidemics in Corrientes had,
and the gruesome failure of not quarantining Corrientes once the mortality
started by December 16, 1870, about two weeks before the deducted beginning of
the outbreak in Buenos Aires.

The same study of inconsistencies between the data and the reconstruction
made us focus on the survival time of those clinically diagnosed
with YF that finally die. The form in which the illness
evolves anticipates the final result. Jones and Wilson \cite{jone72}
indicate the symptoms of cases with a bad prognosis including the
rapidity and degree of jaundice. This information suggests, in terms
of  modeling, that death is not one of two possible outcomes at the
end of the ``toxic period'', as we have first thought.
Separation of to-recover and to-die subpopulations could (should?)
be performed earlier in the development of the illness, each subpopulation
having its own parameters for the illness. Yet, while in theory this
would be desirable, in practice it would have, for the time being,
no effect, since the characteristic periods of the illness have not
been measured in these terms.

Epidemics transmitted by vectors come to an end either when the susceptible
population has been sufficiently exposed so that the replication of
the virus is slowed down (the classical consideration in SIR models)
or when the vector's population is decimated by other (for example,
climatic) reasons. The model shows that both situations can be distinguished
in terms of the mortality statistics.

We have also shown that the total mortality of the epidemic is not
difficult to adjust by changing the death probability of the toxic
phase, and as such, it is not a demanding test for a model. The daily
mortality, when normalized, shows sensitivity to the mosquito abundance,
specially in the evolution times involved, since the general qualitative
shape appears to be fixed. In particular, the date in which the epidemic
reaches half the total mortality is advanced by larger mosquito populations.
However, only comparison of the simulated and historical daily mortality
put enough constraints to the free data in the model (date of arrival
of infected people and mosquito population) to allow for a selection
of possible combinations of their values.

As successful as the model appears to be, it is completely unable
to produce the total mortality in the city, or the spatial extension
of the full epidemic. The simulations produce with BSx4 less than
4500 deaths, while in the historic record, the total mortality in
the city is above 13000 cases. The historical account, and the recorded
data, show that after the initial San Telmo focus has developed, a
second focus in the police district 13 (see Fig. \ref{districtos})
developed, shortly several other foci developed that could not be
tracked \cite{penn95}. Unless the spreading of the illness by infected
humans is introduced (or some other method to make long jumps by the
illness), such events cannot be described. It is worth noticing that
the mobility patterns in 1871 are expected to be drastically different
from present patterns, and as such, the application of models with
human mobility \cite{barm11} is not straightforward and requires
a historical study.

One of the most important conclusions of this work is that the logical
consistency of mathematical  modeling puts a limit to ad-hoc hypotheses,
so often used in a-posteriori explanations, as it forces to accept
not just the desired consequence of the hypotheses, but all other
consequences as well. 

Last, eco-epidemiological models are adjusted to vector populations
pre-existing the actual epidemics and can therefore be used in prevention
to determine epidemic risk and monitor eradication campaigns. In the
present work, the tuning was performed in epidemic data only because
it is actually impossible to know the environmental conditions more
than one hundred years ago. Yet, our wild initial guess for the density
of breeding sites resulted sufficiently close to allow further tuning.

\section*{Acknowledgments}

We want to thank Professor Guillermo Marshall who has been very kind
allowing MLF to take time off her duties to complete this work. We
acknowledge the grant PICTR0087/2002 by the ANPCyT (Argentina) and
the grants X308 and X210 by the Universidad de Buenos Aires. Special
thanks are given to the librarians and  personnel of the Instituto Histórico
de la Ciudad de Buenos Aires, Biblioteca Nacional del Maestro, Museo
Mitre and the library of the School of Medicine UBA.

\bibliography{referencias}
\bibliographystyle{unsrt}

\appendix
\section{Appendix} \label{apendice} 

\subsection{Populations and events of the stochastic transmission model}\label{appendix-subsec1}

We consider a two dimensional space as a mesh of squared patches where the
dynamics of vectors, hosts and the disease take place. Only adult mosquitoes, Flyers, can fly from one
patch to a next one according to a diffusion-like process. 
The coordinates of a patch are given by two indices, $i$ and $j$, corresponding
to the row and column in the mesh. If $X_k$ is a subpopulation in the stage
$k$, then $X_k(i,j)$ is the $X_k$ subpopulation in the patch of coordinates
$(i,j)$.

Population of both hosts (Humans) and vectors ({\em Aedes aegypti}) were
divided into subpopulations representing disease status: SEI for the vectors
and SEIrRTD for the human population.

Ten different subpopulations for the mosquito were taken into account, three
immature subpopulations: eggs $E_{(i,j)}$, larvae $L_{(i,j)}$ and pupae
$P_{(i,j)}$, and seven adult subpopulations: non parous adults
$A1_{(i,j)}$, susceptible flyers $Fs_{(i,j)}$, exposed flyers $Fe_{(i,j)}$,
infectious flyers $Fi_{(i,j)}$ and parous adults in the three
disease status: susceptible $A2s_{(i,j)}$, exposed $A2e_{(i,j)}$ and infectious
$A2i_{(i,j)}$.

The $A1_{(i,j)}$ is always susceptible, after a blood meal it becomes a flyer,
susceptible $Fs_{(i,j)}$ or exposed $Fe_{(i,j)}$, depending on the disease
status of the host. If the host is infectious, $A1_{(i,j)}$ becomes an exposed
flyer $Fe_{(i,j)}$ but if the host is not infectious, then the $A1_{(i,j)}$
becomes a susceptible flyer $Fs_{(i,j)}$. The transmission of the virus depends not only on the
contact between vector and host but also on the transmission probability of the
virus. In this case, we have two transmission probabilities: the transmission probability from 
host to vector $ahv$ and the transmission probability from vector to host $avh$.

Human population $Nh_{(i,j)}$ was split into seven different subpopulations
according to the disease status: susceptible humans $Hs_{(i,j)}$, exposed
humans $He_{(i,j)}$, infectious humans $Hi_{(i,j)}$, humans in remission state $Hr_{(i,j)}$,
toxic humans $Ht_{(i,j)}$, removed humans $HR_{(i,j)}$ and dead humans because of the disease $Hd_{(i,j)}$.

The evolution of the seventeen subpopulations is affected by events that occur
at rates that depend on subpopulation values and some of them also on
temperature, which is a function of time since it changes over the course of
the year seasonally \cite{oter06,oter08}.

\subsection{Events related to immature stages}\label{appendix-subsec2}

Table \ref{events on immatures}  summarizes the events and rates related to
immature stages of the mosquito during their first gonotrophic cycle. The
construction of the transition rates and the election of model parameters
related to the mosquito biology such as: $me$ mortality of eggs, $elr$
hatching rate, $ml$ mortality of larvae, $\alpha$
density-dependent mortality of larvae, $lpr$ pupation rate, $mp$:
mortality of pupae, $par$ pupae into adults development coefficient and the
$ef$ emergence factor were described in detail previously
\cite{oter06,oter08}.

\begin{table*}[hbt]
\begin{center}
\footnotesize
\begin{tabular}{p{3.0cm}p{3.5cm}p{5.0cm}}
\hline\hline Event & Effect & Transition rate\\
\hline

Egg death & $E_{(i,j)} \rightarrow E_{(i,j)}-1$ &  $me \times E_{(i,j)}$\\ 

Egg hatching &  $E_{(i,j)} \rightarrow E_{(i,j)}-1$ $L_{(i,j)} \rightarrow
L_{(i,j)}+1$ & $elr \times E_{(i,j)}$\\ 

Larval death & $L_{(i,j)} \rightarrow L_{(i,j)}-1$ & $ml\times L_{(i,j)} +
\alpha \times L_{(i,j)} \times (L_{(i,j)}-1)$\\ 

Pupation &  $L_{(i,j)} \rightarrow L_{(i,j)}-1$ $P_{(i,j)} \rightarrow
P_{(i,j)}+1$ & $lpr \times L_{(i,j)}$ \\ 

Pupal death & $P_{(i,j)} \rightarrow P_{(i,j)}-1$ & $(mp+
par\times(1-(ef/2))) \times P_{(i,j)}$ \\ 

Adult emergence & $P_{(i,j)} \rightarrow P_{(i,j)}-1$ $A1_{(i,j)}
\rightarrow A1_{(i,j)}+1$ & $par \times (ef/2)\times P_{(i,j)}$\\ \hline\hline

\end{tabular}
\end{center}
\caption{Event type, effects on the populations and transition rates for
the developmental model. The coefficients are
$me$: mortality of eggs;
$elr$: hatching rate;
$ml$:  mortality of larvae;
$\alpha$: density-dependent mortality of larvae;
$lpr$: pupation rate;
$mp$: mortality of pupae;
$par$: pupae into adults development coefficient;
$ef$: emergence factor.
The values of the coefficients are available in subsections \ref{appendix-subsec6} and \ref{appendix-subsec7}.}
\label{events on immatures}
\end{table*}

The natural regulation of {\em Aedes aegypti} populations is due to intra-specific 
competition for food and other resources in the larval stage. 
This regulation was incorporated into the model as a density-dependent 
transition probability which introduces the necessary
nonlinearities that prevent a Malthusian growth of the population. This effect
was incorporated as a nonlinear correction to the temperature dependent larval
mortality.

Then, larval mortality can be written as: $ml L_{(i,j)} + \alpha L_{(i,j)}\times (L_{(i,j)}-1)$ 
where the value of $\alpha$ can be further decomposed as $\alpha = \alpha_0 / \text{BS}_{(i,j)} \label{BS_{(i,j)}}$ 
with $\alpha_0$ being associated with the carrying capacity of one (standardised) breeding site 
and $\text{BS}_{(i,j)}$ being the density of breeding sites in the $(i,j)$ patch \cite{oter06,oter08}.

\subsection{Events related to the adult stage}\label{appendix-subsec3}

{\em Aedes aegypti} females ($A1$ and $A2$) require blood to complete their gonotrophic cycles. 
In this process, the female may ingest viruses with the blood meal from an infectious 
human during the human Viremic Period $VP$. The viruses develop within the mosquito during the
Extrinsic Incubation Period $EIP$ and then are reinjected into the blood
stream of a new susceptible human with the saliva of the mosquito in later
blood meals. The virus in the exposed human develops during the Intrinsic
incubation Period $IIP$ and then begin to circulate in the blood stream
(Viremic Period),  the human becoming infectious. The flow from susceptible
to exposed subpopulations (in the vector and the host) depends not only on the
contact between vector and host but also on the transmission probability of the
virus. In our case, there are two transmission probabilities: the transmission
probability from host to vector $ahv$ and the transmission probability from
vector to host $avh$.

\begin{table*}[hbt]
\centering
\footnotesize
\begin{tabular}{p{3.0cm}p{3.5cm}p{5.0cm}}
\hline\hline Event & Effect & Transition rate\\
\hline

Adults 1 Death & $A1_{(i,j)} \rightarrow A1_{(i,j)}-1$ & $ma\times A1_{(i,j)}$\\ \vspace{0.005cm} \\

I Gonotrophic cycle with virus exposure & $A1_{(i,j)}\rightarrow
A1_{(i,j)}-1$ $Fe_{(i,j)}\rightarrow Fe_{(i,j)}+1$  &  $cycle1
\times A1_{(i,j)}\times (Hi_{(i,j)}/Nh_{(i,j)})\times ahv$\\ \vspace{0.005cm} \\

I Gonotrophic cycle without virus exposure & $A1_{(i,j)}\rightarrow
A1_{(i,j)}-1$ $Fs_{(i,j)}\rightarrow Fs_{(i,j)}+1$  & $cycle1
\times A1_{(i,j)}\times
((((Nh_{(i,j)}-Hi_{(i,j)})/Nh_{(i,j)})+(1-ahv)\times(Hi_{(i,j)}/Nh_{(i,j)}))$\\\vspace{0.005cm} \\

Oviposition of susceptible flyers &  $E_{(i,j)}\rightarrow E_{(i,j)}+egn$
$Fs_{(i,j)}\rightarrow Fs_{(i,j)}-1$ $A2s_{(i,j)}\rightarrow A2s_{(i,j)}+1$ &
$ovr_{(i,j)}\times Fs_{(i,j)}$\\ \vspace{0.005cm} \\

Oviposition of exposed flyers &  $E_{(i,j)}\rightarrow E_{(i,j)}+egn$ 
 $Fe_{(i,j)}\rightarrow Fe_{(i,j)}-1$ $A2e_{(i,j)}\rightarrow A2e_{(i,j)}+1$
& $ovr_{(i,j)}\times Fe_{(i,j)}$\\ \vspace{0.005cm} \\

Oviposition of infected flyers &  $E_{(i,j)}\rightarrow E_{(i,j)}+egn$ 
$Fi_{(i,j)}\rightarrow Fi_{(i,j)}-1$  $A2i_{(i,j)}\rightarrow A2i_{(i,j)}+1$ &
$ovr_{(i,j)}\times Fi_{(i,j)}$\\ \hline\hline

\end{tabular}
\caption{Event type, effects on the subpopulations and transition rates for
the developmental model. The coefficients are
$ma$: mortality of adults;
$cycle1$: gonotrophic cycle coefficient (number of daily cycles)
for adult females in stages $A1$.;
$ahv$: transmission probability from host to vector;
$ovr_{(i,j)}$: oviposition rate by flyers in the (i,j) patch;
$egn$: average number of eggs laid in an oviposition.
The values of the coefficients are available in Table \ref{parbo}, subsections \ref{appendix-subsec6}, 
\ref{appendix-subsec7}, \ref{appendix-subsec8} and \ref{appendix-subsec9}.}
\label{events on I gonotrophic cycle and oviposition}
\end{table*}

The events related to the adult stage are shown in Table \ref{events on I
gonotrophic cycle and oviposition} to \ref{events on adult mortality}.  
Table \ref{events on I gonotrophic cycle and oviposition}  summarizes 
the events and rates related to adults during their first gonotrophic cycle and related to
oviposition by flyers according to their disease status. 

\begin{table*}[hbt]
\centering
\footnotesize
\begin{tabular}{p{3.0cm}p{3.5cm}p{5.0cm}}
\hline\hline Event & Effect & Transition rate\\
\hline

II Gonotrophic cycle of susceptible Adults 2 with virus exposure &
$A2s_{(i,j)}\rightarrow A2s_{(i,j)}-1$ $Fe_{(i,j)}\rightarrow
Fe_{(i,j)}+1$ & $cycle2 \times A2s_{(i,j)}\times(Hi_{(i,j)}/Nh_{(i,j)})\times ahv$\\
\vspace{0.005cm} \\

II Gonotrophic cycle of susceptible Adults 2 without virus exposure &
$A2s_{(i,j)}\rightarrow A1_{(i,j)}-1$  $Fs_{(i,j)}\rightarrow Fs_{(i,j)}+1$ &
$cycle2
\times A2s_{(i,j)}\times((((Nh_{(i,j)}-Hi_{(i,j)})/Nh_{(i,j)})+(1-ahv)\times(Hi_{(i,j)}/Nh_{(i,j)}))$\\
\vspace{0.005cm} \\

II Gonotrophic cycle of exposed Adults 2 & $A2e_{(i,j)}\rightarrow
A2e_{(i,j)}-1$ $Fe_{(i,j)}\rightarrow Fe_{(i,j)}+1$ & $cycle2
\times A2e_{(i,j)}$\\ \hline\hline

\end{tabular}
\caption{Event type, effects on the subpopulations and transition rates for
the developmental model. The coefficients are
$cycle2$: gonotrophic cycle coefficient (number of daily cycles)
for adult females in stages $A2$.;
$ahv$: transmission probability from host to vector.
The values of the coefficients are available in Table \ref{parbo}, subsections \ref{appendix-subsec6}, 
\ref{appendix-subsec7}, \ref{appendix-subsec8} and \ref{appendix-subsec9}.}
\label{events on II gonotrophic cycle and human contagion A}

\end{table*}

Table \ref{events on II gonotrophic cycle and human contagion A} and Table \ref{events on II
gonotrophic cycle and human contagion B}  summarize the events and rates
related to adult 2 gonotrophic cycles, exposed Adults 2 and exposed flyers
becoming infectious and human contagion.

\begin{table*}[hbt]
\centering
\footnotesize
\begin{tabular}{p{3.0cm}p{3.4cm}p{4.5cm}}
\hline\hline Event & Effect & Transition rate\\
\hline

Exposed Adults 2 becoming infectious & $A2e_{(i,j)}\rightarrow
A2e_{(i,j)}-1$ $A2i_{(i,j)}\rightarrow A2i_{(i,j)}+1$ & $(1/(EIP-(1/ovr_{(i,j)})))A2e_{(i,j)}$\\ 
\vspace{0.005cm} \\
Exposed flyers becoming infectious & $Fe_{(i,j)}\rightarrow Fe_{(i,j)}-1$
$Fi_{(i,j)}\rightarrow Fi_{(i,j)}+1$ & $(1/(EIP-(1/ovr_{(i,j)})))Fe_{(i,j)}$\\ \vspace{0.005cm} \\

II Gonotrophic cycle of infectious Adults 2 without human contagion &
$A2i_{(i,j)}\rightarrow A2i_{(i,j)}-1$ $Fi_{(i,j)}\rightarrow Fi_{(i,j)}+1$
$Hs_{(i,j)}\rightarrow Hs_{(i,j)}-1$ $He_{(i,j)}\rightarrow He_{(i,j)}+1$ &
$cycle2 \times A2i_{(i,j)}\times(Hs_{(i,j)}/Nh_{(i,j)})\times avh$\\ \vspace{0.005cm} \\

II Gonotrophic cycle of infectious Adults 2 without human contagion &
$A2i_{(i,j)}\rightarrow A2i_{(i,j)}-1$ $Fi_{(i,j)}\rightarrow Fi_{(i,j)}+1$ &
$cycle2
\times A2i_{(i,j)}\times((((Nh_{(i,j)}-Hs_{(i,j)})/Nh_{(i,j)})+(1-avh)\times(Hs_{(i,j)}/Nh_{(i,j)}))$\\
\hline\hline

\end{tabular}
\caption{Event type, effects on the subpopulations and transition rates for
the developmental model. The coefficients are
$cycle2$: gonotrophic cycle coefficient (number of daily cycles)
for adult females in stages $A2$;
$ovr_{(i,j)}$: oviposition rate by flyers in the $(i,j)$ patch;
$avh$: transmission probability from vector to host;
$EIP$: extrinsic incubation period.
The values of the coefficients are available in Table \ref{parbo}, 
subsections \ref{appendix-subsec6}, \ref{appendix-subsec7}, \ref{appendix-subsec8} and \ref{appendix-subsec9}}
\label{events on II gonotrophic cycle and human contagion B}

\end{table*}

Table \ref{events on adult mortality}  summarizes the events and rates related to non parous adult (Adult 2) and Flyer
death.

\begin{table*}[hbt]
 \centering
\footnotesize
\begin{tabular}{ccc}
\hline\hline Event & Effect & Transition rate\\
\hline

Susceptible flyer Death & $Fs_{(i,j)} \rightarrow Fs_{(i,j)}-1$ & $ma \times Fs_{(i,j)}$\\ 

Exposed flyer Death & $Fe_{(i,j)} \rightarrow Fe_{(i,j)}-1$ & $ma\times Fe_{(i,j)}$\\ 

Infectious flyer Death & $Fi_{(i,j)} \rightarrow Fi_{(i,j)}-1$ & $ma\times Fi_{(i,j)}$\\ 

Susceptible Adult 2 Death & $A2s_{(i,j)} \rightarrow A2s_{(i,j)}-1$ & $ma\times A2s_{(i,j)}$\\ 

Exposed Adult 2 Death & $A2e_{(i,j)} \rightarrow A2e_{(i,j)}-1$ & $ma\times A2e_{(i,j)}$\\ 

Infectious Adult 2 Death & $A2i_{(i,j)} \rightarrow A2i_{(i,j)}-1$ & $ma\times A2i_{(i,j)}$\\ \hline\hline

\end{tabular}
\caption{Event type, effects on the subpopulations and transition rates for
the developmental model. The coefficients are $ma$: adult mortality.
The values of the coefficients are available in subsection \ref{appendix-subsec7}}
\label{events on adult mortality}

\end{table*}

\subsection{Events related to flyer dispersal}\label{appendix-subsec4}

Some experimental results and observational studies show
that the {\em Aedes aegypti} dispersal is driven by
the availability of oviposition sites \cite{wolf53,reit95,edma98}. According
to these observations, we considered that only the Flyers $F_{(i,j)}$ can fly
from patch to patch in search of oviposition sites. The implementation of flyer
dispersal has been described elsewhere \cite{oter08}.

The general rate of the dispersal event is given by: $\beta \times F_{(i,j)}$, where $\beta$ 
is the dispersal coefficient and $F_{(i,j)}$ is the Flyer population which can be susceptible
$Fs_{(i,j)}$, exposed $Fe_{(i,j)}$ or infectious $Fi_{(i,j)}$ depending on the
disease status.

The dispersal coefficient $\beta$ can be written as

\begin{equation} \beta = \left\{
\begin{array}{cc}
0 & \mbox{\footnotesize if the patches are disjoint}\\
diff/d_{ij}^2 & \mbox{\footnotesize if the patches have}\\
& \mbox{\footnotesize at least a common point}\label{migration}
\end{array} \right.
\end{equation}

where $d_{ij}$ is the distance between the centres of the patches and $diff$ is a diffusion-like coefficient so that dispersal is compatible with a
diffusion-like process \cite{oter08}.

\subsection{Events related to human population}\label{appendix-subsec5}

\begin{table*}[hbt]
\centering
\footnotesize
\begin{tabular}{p{4.5cm}p{4.5cm}p{4.0cm}}
\hline\hline Event & Effect & Transition rate\\
\hline

Born of susceptible humans & $Hs_{(i,j)} \rightarrow Hs_{(i,j)}+1$ &
$mh\times Nh_{(i,j)}$\\ \vspace{0.005cm} \\

Death of susceptible humans & $Hs_{(i,j)} \rightarrow Hs_{(i,j)}-1$ &
$mh\times Hs_{(i,j)}$\\ \vspace{0.005cm} \\

Death of exposed humans & $He_{(i,j)} \rightarrow He_{(i,j)}-1$ & 
$mh\times He_{(i,j)}$\\ \vspace{0.005cm} \\

Transition from exposed to viraemic & $He_{(i,j)} \rightarrow
He_{(i,j)}-1$ $Hi_{(i,j)}\rightarrow Hi_{(i,j)}+1$ & $(1/IIP)\times He_{(i,j)}$\\ \vspace{0.005cm} \\

Death of Infectious humans & $Hi_{(i,j)} \rightarrow Hi_{(i,j)}-1$ &
$mh\times Hi_{(i,j)}$\\ \vspace{0.005cm} \\

Transition from infectious humans to humans in remission state& $Hi_{(i,j)} \rightarrow Hi_{(i,j)}-1$ 
$Hr_{(i,j)}\rightarrow Hr_{(i,j)}+1$ & $(1/VP)\times Hi_{(i,j)}$\\ 

Death of humans in remission state& $Hr_{(i,j)} \rightarrow Hr_{(i,j)}-1$ & $mh\times Hr_{(i,j)}$\\ \hline

Transition from humans in remission to toxic humans& $Hr_{(i,j)} \rightarrow Hr_{(i,j)}-1$ 
$Ht_{(i,j)}\rightarrow Ht_{(i,j)}+1$ & $((1-rar)/rP)\times Hr_{(i,j)}$\\ \vspace{0.005cm} \\

Recovery of humans in remission& $Hr_{(i,j)} \rightarrow Hr_{(i,j)}-1$ 
$HR_{(i,j)}\rightarrow HR_{(i,j)}+1$ & $(rar/rP)\times Hr_{(i,j)}$\\ 

Death of removed humans & $HR_{(i,j)} \rightarrow HR_{(i,j)}-1$ & $mh\times HR_{(i,j)}$\\ \hline

Death of toxic humans & $Ht_{(i,j)} \rightarrow Ht_{(i,j)}-1$ 
$Hd_{(i,j)}\rightarrow Hd_{(i,j)}+1$ & $(mt/tP)\times Ht_{(i,j)}$\\ \vspace{0.005cm} \\

Recovery of toxic humans& $Ht_{(i,j)} \rightarrow Ht_{(i,j)}-1$ 
$HR_{(i,j)}\rightarrow HR_{(i,j)}+1$ & $((1-mt)/tP)\times Ht_{(i,j)}$\\ \hline \hline

\end{tabular}
\caption{Event type, effects on the subpopulations and transition rates for
the developmental model. The coefficients are
$mh$: human mortality coefficient;
$VP$: human viremic period;
$mh$: human mortality coefficient;
$IIP$: intrinsic incubation period;
$rP$: remission period;
$tP$: toxic period;
$rar$: recovery after remission probability;
$mt$: mortality probability for toxic patients.
The values of the coefficients are available in Table \ref{parbo}.}
\label{events on humans}

\end{table*}

Human contagion has been already described in Table \ref{events
on II gonotrophic cycle and human contagion B}.
Table \ref{events on humans}  summarizes the events and rates in which humans
are involved. The human population was fluctuating but balanced, meaning that the birth
coefficient was considered equal to the mortality coefficient $mh$.

\subsection{Developmental Rate coefficients}\label{appendix-subsec6}

The developmental rates that correspond to egg hatching, pupation, adult
emergence and the gonotrophic cycles were evaluated using the results of the
thermodynamic model developed by Sharp and DeMichele \cite{shar77} and
simplified by Schoofield \emph{et al.} \cite{scho81}. According to this model, the
maturation process is controlled by one enzyme which is active in a given
temperature range and is deactivated only at high temperatures. The development
is stochastic in nature and is controlled by a Poisson process with rate
$R_D(T)$. In general terms, $R_D(T)$ takes the form

\begin{align}
\label{enzyme1}
R&_D(T) = R_D(298\text{ K}) \\
&\times \frac{(T/298\text{ K})
\exp((\Delta H_A/R) (1/298\text{ K}-1/T))}
{1 + \exp(\Delta H_H/R)(1/T_{1/2}-1/T))} \notag 
\end{align}
where $T$ is the absolute temperature, $\Delta H_A$ and $\Delta H_H$ are
thermodynamics enthalpies characteristic of the organism, $R$ is the universal
gas constant, and $T_{1/2}$ is the temperature when half of the enzyme is
deactivated because of high temperature.

Table \ref{entalpias} presents the values of the different coefficients
involved in the events: egg hatching, pupation, adult emergence and
gonotrophic cycles. The values are taken from Ref. \cite{fock93a} and are
discussed in Ref. \cite{oter06}.

\begin{table*}[hbt]
 \centering
\begin{tabular}{c|c|c|c|c|c}
\hline \hline
  Develop. Cycle (\ref{enzyme1})& $R_D(T)$ & $R_D(298\text{ K})$ & $\Delta H_A$
  & $\Delta H_H$ & $T_{1/2}$ \\
 \hline 
 Egg hatching & elr & 0.24 & 10798 & 100000 & 14184 \\
 Larval develop. & lpr & 0.2088 & 26018 & 55990 & 304.6 \\
 Pupal Develop.  & par  & 0.384 & 14931 & -472379 & 148 \\
 Gonotrophic c. ($A1$) & cycle1 & 0.216 & 15725 & 1756481 & 447.2 \\
 Gonotrophic c. ($A2$) & cycle2 & 0.372 & 15725 & 1756481 & 447.2 \\ \hline \hline
\end{tabular}
\caption{Coefficients for the enzymatic model of maturation [Eq. (\ref{enzyme1})]. $R_D$ is measured 
in day$^{-1}$, enthalpies are measured in (cal / mol) and the temperature $T$ is measured in absolute (Kelvin) degrees.
\label{entalpias}}
\end{table*}

\subsection{Mortality coefficients}\label{appendix-subsec7}

\paragraph*{Egg mortality.} The mortality coefficient of eggs is $me =
0.01$ 1/day, independent of temperature in the range $278\text{ K} \le T \le
303\text{ K}$ \cite{trpi72}. 

\paragraph*{Larval mortality.} The value of $\alpha_0$ (associated to the
carrying capacity of a single breeding site) is $\alpha_0 = 1.5$,  and was
assigned by fitting the model to observed values of immatures in the cemeteries
of Buenos Aires \cite{oter06}. The temperature dependent larval death
coefficient is approximated by $ml=0.01 + 0.9725 \exp(-(T-278)/2.7035)$
and it is valid in the range $278\text{ K} \le T \le  303\text{ K}$ \cite{hors55,bar-58,rued90}. 

\paragraph*{Pupal mortality.} The intrinsic mortality of a pupa has been
considered as $mp= 0.01 + 0.9725 exp(-(T-278)/2.7035)$ \cite{hors55,bar-58,rued90}. 
Besides the daily mortality in the pupal stage, there is an additional mortality contribution associated 
to the emergence of the adults. We considered a mortality of $17\%$ of the pupae at this 
event, which is added to the mortality rate of pupae. Hence, the emergence factor is $ef = 0.83$
\cite{sout72}. 

\paragraph*{Adult mortality.} Adult mortality coefficient is $ma=0.09
1/\text{day}$ and it is considered independent of temperature in the range $278\text{ K}
\le T \le  303\text{ K}$ \cite{hors55,chri60,fayr64}.

\subsection{Fecundity and oviposition coefficient}\label{appendix-subsec8}

Females lay a number of eggs that is roughly proportional to their body weight
($46.5 \text{ eggs/mg}$) \cite{bar-57,naya75}. Considering that the mean weight of
a three-day-old female is $1.35\text{ mg}$ \cite{chri60}, we estimated the average
number of eggs laid in one oviposition as $egn= 63$.

The oviposition coefficient $ovr_{(i,j)}$ depends on breeding site density $\text{BS}_{(i,j)}$
and it is defined as: \begin{equation}ovr_{(i,j)}= \left\{
\begin{array}{cccc} &\theta/tdep&\mbox{ if } & \text{BS}_{(i,j)} \leq 150 \\
&1/tdep&\mbox{ if } & \text{BS}_{(i,j)} > 150 \end{array}\right.
\label{oviposition2}\end{equation}

where $\theta$ was chosen as  $\theta=\text{BS}_{(i,j)}/150$, a linear function of the density
of breeding sites \cite{oter08}.

\subsection{Dispersal coefficient}\label{appendix-subsec9}

We chose a diffusion-like coefficient of $diff$ = 830 m$^2$/day which
corresponds to a short dispersal, approximately a mean dispersal of 30 m in one
day, in agreement with short dispersal experiments and field studies  analyzed
in detail in our previous article \cite{oter08}.

\subsection{Mathematical description of the stochastic model}\label{appendix-subsec10}

The evolution of the subpopulations is  modeled by a state dependent Poisson
process \cite{ethi86,ande00} where the probability of the state:

\begin{align*}
  (E, L, P, &A1, A2s, A2e, A2i, Fs, Fe, Fi, \\ 
  &Hs, He, Hi, Hr, Ht, HR, Hd)_{(i,j)}
\end{align*}
evolves in time following a Kolmogorov forward equation that can be constructed
directly from the information collected in Tables  \ref{events on immatures} to
\ref{events on humans} and in Eq. \ref{migration}.

\subsection{Deterministic rates approximation for the density-dependent
Markov process}\label{appendix-subsec11}

Let $X$ be an integer vector having as entries the populations under
consideration, and $e_\alpha, \alpha=1\dots \kappa$ the events at which the
populations change by a fixed amount $\Delta_\alpha$ in a Poisson process with
density-dependent rates. Then, a theorem by Kurtz \cite{ethi86} allows us to
rewrite the stochastic process as: \begin{equation} X(t) = X(0) +
\sum_{\alpha=1}^\kappa \Delta_\alpha Y(\int_{0}^t\omega_\alpha(X(s))ds)
\label{kurtz} \end{equation} where $\omega_\alpha(X(s)$ is the transition rate
associated with the event $\alpha$ and $Y(x)$ is a random Poisson process of
rate $x$.

The deterministic rates approximation to the stochastic process represented by
Eq. (\ref{kurtz}) consists  of the introduction of a deterministic approximation for
the arguments of the Poisson variables $Y(x)$ in Eq. (\ref{kurtz})
\cite{sola03,apar01b}. The reasons for such a proposal is that the transition
rates change at a slower rate than the populations. The number of each kind of
event is then approximated as independent Poisson processes with deterministic
arguments satisfying a differential equation.

The probability of $n_\alpha$ events of type $\alpha$ having occurred after a
time $dt$ is approximated by a Poisson distribution with parameter
$\lambda_\alpha$. Hence, the probability of the population taking the value
 
\begin{equation}
X=X_0 + \sum_{\alpha=1}^\kappa \Delta_\alpha n_\alpha
\end{equation}
at a time interval $dt$ after being in the state $X_0$ is approximated by a
product of independent Poisson distributions of the form

\begin{equation}
\mbox{Probability}(n_1\dots n_\kappa, dt / X_0) =
\prod_{\alpha=1}^\kappa P_\alpha(\lambda_\alpha)
\end{equation}
and

\begin{equation}
P^\alpha_{n_1\dots n_E}(\lambda_\alpha) = \exp(-\lambda_\alpha) \frac{\lambda_\alpha^{n_\alpha}}{n_\alpha!}
\end{equation}
whenever $X=X_0 + \sum_{\alpha=1}^\kappa \Delta_\alpha n_\alpha$ has no negative
entries and 

\begin{align}
P^\alpha_{n_1\dots n_E}(\lambda_\alpha) &= \exp(-\lambda_\alpha) \sum_{i=n_\alpha}^\infty 
\frac{\lambda_\alpha^{i}}{i!} \notag \\
&= 1-\exp(-\lambda_\alpha) \sum_{i=0}^{n_\alpha-1}\frac{\lambda_\alpha^{i}}{i!}
\end{align}
if $\{n_i\}$ makes a component in $X$ zero (see Ref. \cite{sola03})

Finally,
\begin{equation}
d\lambda_\alpha/dt =  <\omega_\alpha(X)>
\end{equation}
where the averages are taken self-consistently with the proposed
distribution ($\lambda_\alpha(0)=0$).

The use of the Poisson approximation represents a substantial saving of
computer time compared to direct (Monte Carlo) implementations of the
stochastic process.

\end{document}